\def\beq{\begin{equation}}
\def\eeq{\end{equation}}
\def\baq{\begin{eqnarray}}
\def\eaq{\end{eqnarray}}

\def\figloc#1#2{\epsfysize=3in
    \centerline{\epsfbox{fig#1.ps}}
    \centerline{Figure #1}
    {\raggedright\it   #2 } 
    \bigskip
    }

\documentclass[preprint,aps,tightenlines]{revtex4}

\begin{document}
\input epsf
\title{ Large Universe as the initial condition for a Collapsing Universe}

\author{ Daniel Green and W. G. Unruh}
\affiliation{ CIAR Cosmology and Gravity Program\\
Dept. of Physics and Astronomy\\
University of B. C.\\
Vancouver, Canada V6T 1Z1\\
~
email: unruh@physics.ubc.ca}

~

~

\begin{abstract}
Using the extension of the standard Hawking-Hartle prescription for defining a
wave function for the universe, we show that it is possible, given a suitable
form for the scalar field potential, to have the universe begin at its largest
size and thereafter contract, with the growth of perturbations proceeding from
small, at the largest size, to largest when the universe is small. This solution
would dominate the wave function by an exponentially large amount if one
chooses the Hartle Hawking prescription for the wave-function, but is
exponentially sub-dominant for the Linde-Vilenkin prescription.

\end{abstract}
\maketitle

\section{Large Initial Universe}
In a surprising result, Hartle and Hawking twenty years ago showed \cite{HH} how, at
least in the semi-classical theory of quantum gravity, one could give an
intuitively appealing definition for a preferred wave function for the universe.
If one believes that nature selects this definition, then they argued that this
wave function explains  various features which we observe the universe to have.
One of these features is that the universe is simple when it is small \cite{HHall} and
becomes more and more complex as it grows in size. While Hawking originally
believed that the universe must always be simple when small \cite{Haw}, Page and 
Laflamme \cite{Page} pointed out that this wave-function would be expected to also contain
probablilities for the universe to be complex when small. This was interpreted
as indicating a classical history of the universe which began simple when small,
expanded, with the gravitational instability causing growth in the quantum
fluctuations present when small. The universe would eventually re-collapse with
the fluctuations continuing to grow. If one wrote the HH wave-function in terms
of a basis which contained such semi-classical components, one would interpret
them to say that the thermodynamic arrow of time would be driven by these
fluctuations and that one should interpret those solutions such that the low
fluctuation end should be interpreted as the beginning and the high as the end
of the evolution.

However, there is an alternative interpretation of the HH wave-function(
see for example Jheeta and Unruh\cite{JU}). The
prescription bears a strong resemblance to the prescription of a semi-classical
WKB wave-function near a classical turning point of a potential. The classical
solution for the HH models is that of a De Sitter space with the the cosmological
constant dominating the evolution-- ie the classical solution looks like
De Sitter space in spherical coordinates near the point where the spherical space
reaches it minimum size. 

Interpreted in this way there is another place where the universe could and does
bounce; namely if a closed universe has matter such that the energy density of
the matter drops as the universe expands. In this case, the curvature terms in
the FRW equations eventually dominates and halts the expansion and the universe
re-collapses. The question which we address is whether the HH prescription
could be used to place boundary conditions at this bounce point rather than at
the point where the universe achieves it minimum size?

We will examine a simple model, with a homogeneous scalar field driving the
dynamics of the universe. The space-time will be assumed to be a closed spherical
homogeneous and isotropic FRW cosmology with metric

\beq
ds^2= N^2 dt^2 - a(t)^2 \left(dr^2 +sin(r)^2 \left(d\theta^2+ sin(\theta)^2
d\phi^2\right)\right)
\eeq

The mini-superspace action for this space-time becomes
\beq
S= \int N a^3 \left(-6{\dot a^2\over  N^2 a^2} + {6\over a^2} + {1\over 2N^2}\dot\phi ^2
-V(\phi) \right)dt 
\eeq
which has as semi-classical equations of motion

\baq
{d\over dt} {a^3\over N} {d\phi\over dt} + Na^3{dV(\phi)\over d\phi}=0\\
12a{d\over Ndt} {da\over Ndt}+6 {da\over Ndt}^2 +6 +3a^2({1\over 2N^2} \dot\phi^2
- V)=0\\
\left({\dot a ^2\over 2 N^2 a^2} +{6\over a^2} - {1\over 2N^2}\dot\phi ^2
-V(\phi) \right)=0
\eaq
where $\dot{~}={d\over dt}$.
(The third is the constraint equation, and is an integral of motion
 of the other two).

The semi-classical HH prescription is that the path integral will be 
dominated by
a solution of the equations which has only a final boundary given by the
classical state on which one wishes to evaluate the wave function
,(I.e., the single boundary is defined by the values of $a$ and $\phi$ (or
$\dot\phi$) which one wants on the final boundary). We must find solutions to these
equations which have no initial boundary-- which have $a(t)$ going to zero and
$\phi(t)$ regular as $a(t)$ approaches zero. This is impossible if $N$ ( and $a$) are
real. We can however find such solutions if $N$ is chosen to be complex, and in
particular if we choose $N$ to become purely imaginary as $a$ approaches
zero. (For a generalisation to arbitrary complex $N$ see Jheeta and
Unruh \cite{JU}). 
One  choice, if it works,  is to take $N$ and $a$ to be  purely real over some 
range of times (the Lorentzian regime), and then where $a(t)$
has zero derivative, to take $N$ to be purely imaginary. This allows one to
smoothly join $a(t)$ across the surface where $N$ discontinuously.

However, in general, both $\dot a$ and $\dot\phi$ cannot be taken to be zero at the same
time $t$ and still find a solution with only a single boundary. 
Since the equations of motion for $a$ and $\phi$ are smooth second order
equations, the derivatives $d\over N dt$ must also be smooth. While we can keep
$a(t)$ real across this boundary, since going from real zero to imaginary zero is
a continuous transformation, we cannot do the same for $\phi$. Thus, $\phi$ must thus
become complex as one crosses this boundary. 

We are going to try to model the universe such that at time $t=0$ we have
$\dot a=0$, and for $t>0$, the universe contracts from the value of $a$ at this
time. Ie, we take $t=0$ to be the time at which the universe achieves its
maximum value in the Lorentzian regime.

Examining the constraint equation for $a$, let us examine this boundary where
$a(t)=0$. Since $\dot a$ is, by assumption less, than zero (the universe is
contracting) for larger values of $t$,  the potential term must be
sub-dominant.  The curvature term $6a$ cancels with the kinetic 
energy. If $V$ were to dominate, then $a^2 V$ decreases as $a$ decreases,
and ${\dot a}^2$ would become imaginary. 
 Thus, in order that we have such a maximum in the radius of the universe,
  the term $a^3\dot\phi^2$ must be dominating the
energy, and $\dot\phi^2$ must be decreasing at least as fast as $1/a^2$ But for
very small $V$ this is precisely how $\dot \phi$ is doing. (For small ${dV\over
d\phi}$, 
$\phi$ obeys
\beq
{d\over dt} {a^3} {d\phi\over Ndt}\approx 0
\eeq
or 
\beq
{d\phi\over Ndt}\approx {1\over a^3}
\eeq
Ie, in order that the universe re-collapse, the potential term cannot dominate
the evolution of the universe-- it must be dominated by the kinetic energy of
the matter. 

Now, ${1\over N } \dot\phi$ must be continuous across that boundary where $N$
goes from real to imaginary. Since we
are taking N to be purely imaginary for $t<0$,  $\dot \phi$ must also  be
imaginary for times just before zero. Ie, $\phi$ must pick up
an imaginary part. 

Let us now examine the evolution of this model in this realm where $N$ is purely
imaginary. Let us assume that the potential $V(\phi)$ is such that the potential
remains real for $\phi=\phi_0 +i\Phi$, where $\phi_0$ is the value of $\phi$ at
the point where $\dot a=0$. Since $N$ is arbitrary, we will take $N=1$ for $t>0$
and $N=i$ for $t<0$.  The equation of motion
for $\Phi$ will be 
\beq
{d\over d\tau} a^3 {d\Phi\over dt} = {d V\over d\Phi}
\eeq
and for $a$, the constraint equation gives
\beq
\left({da\over dt}\right)^2 = -1  + {1\over 6}a^2({1\over 2}{d\Phi\over dt}^2 + 
V(\Phi))
\eeq

The second order Einstein equation for $a$ shows that through $t=0$, $\ddot a$
reverses sign. Thus, while the universe contracts toward the future in the
Lorentzian regime, it must expand into the past at this turning point.

 Thus 
$da\over dt$ is less than zero for $t<0$ near 0, and the universe expands as go
to negative $t$.  This raises the question of how
one could get $a$ to go to zero in this N-imaginary region. The kinetic energy
which dominates is driving the universe larger.  The answer is that
as $a$ increases, $\dot\Phi$ keeps decreasing. If the potential $V(\Phi)$
remains non-zero, eventually it will dominate once again. Once it dominates, the
potential term will keep increasing until it becomes of the same order as the
curvature term (the 1 in the equation for $\dot a$) and $a$ has another turning
point. The universe will then collapse as a function of $t$, the kinetic term
will again grow, and the danger is that one will hit another turning point. If
however the potential is arranged so that during this collapse, the potential
increases, and thus $\dot \Phi$ decreases, one can arrange the system so that
$a$ can collapse all the way to zero smoothly.

Let us examine a specific model. We take the potential to be 
\beq
V(\phi)= V_0 +\alpha \phi^4.
\eeq
We are going to assume that at t=0, $\phi(0)=\phi_0=0$, $a(0) $ is large, and 
$ \dot a(0)=0$. We will assume that $V_0$ is  smaller than
$1/a(0)^2 $ and thus  $\dot \phi^2 \approx {12\over a(0)^2}$. 

If the solution is such that $a$ approaches zero smoothly at $t=t_0$
(so that $\Phi$ and
$\dot\Phi$ are both finite there) then near $t=t_a$ we have

\baq
\Phi(t)\approx \Phi(t_0)( 1-{\alpha\over 8}(t-t_0)^2+...)
\\
a(t)=(t-t_0)(1- (V_0-{\alpha\over 4}\Phi(t_0)^4)(t-t_0)^2
+...)
\eaq
 
\figloc{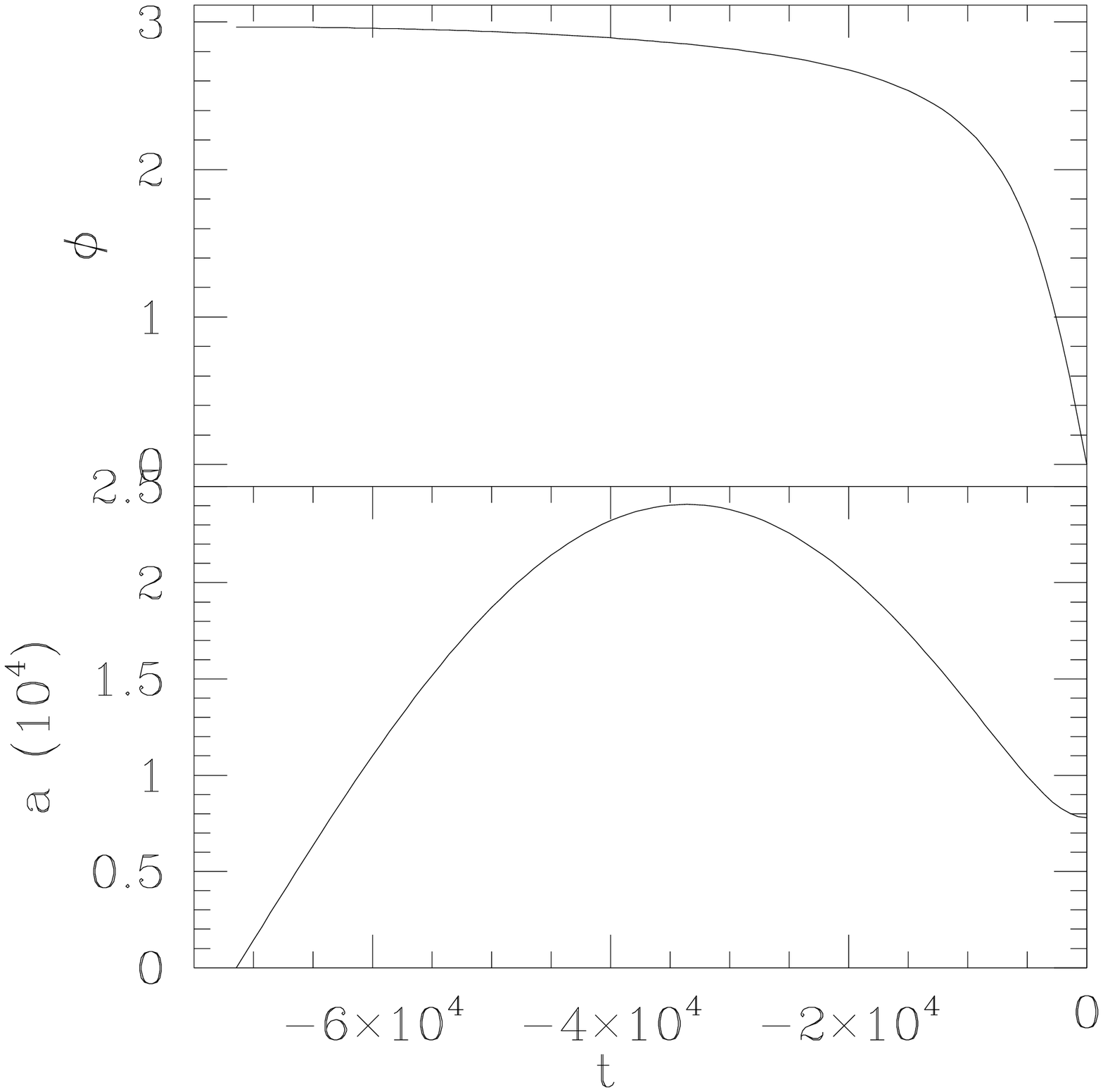}{The evolution of $\phi$ and $a$ in time in the $N=i$ regime. We have
 $V_0=10^-8$, $\alpha$ the quartic coefficient is $2~10^{-11}$. The  value
 of $a$ at $t=0$ is carefully tuned. }
 
 In figure 1 we plot a solution with $V_0= 10^{-8}$ and $\alpha=2~10^{-11}$. The
 lower axis is
 Note that the field $\Phi$ approaches a constant for most of the evolution.

 If the potential in the imaginary direction is sufficiently flat for a large enough 
 value of $\Phi$, the universe can oscillate between the the
 kinetic $\dot\Phi$ dominated bounce at smaller values of $a$ and the potential
 dominated one at large values of $a$ for  many bounces . In Figure 2 we have a
 solution with $V_0=10^-8$ and $\alpha= 10^{-13}$, where we have three
 bounces before the radius eventually goes to zero. 
 
 In the $N$ real regime, the potential $a$ will decrease, and the kinetic
 dominated term will grow roughly as $\dot\phi= 1/a^3$. If the kinetic energy
 dominates, the solution for $a$ goes as $t^1/3$, and $\phi$ grows
 logarithmically. If the potential in the real $\phi$ direction grows, it may
 eventually dominate the matter again, and the universe would suffer a De Sitter
 like bounce at small values of $a$.
 
 Figure 2 is the solution for $N$ and $\phi$ real into the past from the large
 radius bounce which matches to the solution for Figure 1. In this particular
 case, because the potential eventually grows so rapidly with $\phi$, the
 smaller bounce radius is not that much smaller than the largest size. However
 this depends entirely on the behaviour of the potential for large real $\phi$.
 Thus, if our potential were say $V_0 \cos(\phi(t)/\phi_a)$, with $\phi_a$ very
 large, our potential along the real $\phi$ axis would remain bounded by
 $\phi_0$, and the dynamics of $a$ would be kinetic dominated for all times to
 the past of the large bounce. In the imaginary direction, $V(\phi)$ would have
 the required confining shape and would, once $\phi$ had grown to of order
 $\phi_a$, allow for the radius to go smoothly to zero. In the real $\phi$ and
 $N$ regime, the solution would be kinetic dominated all the way  to a
  singularity.
  \figloc{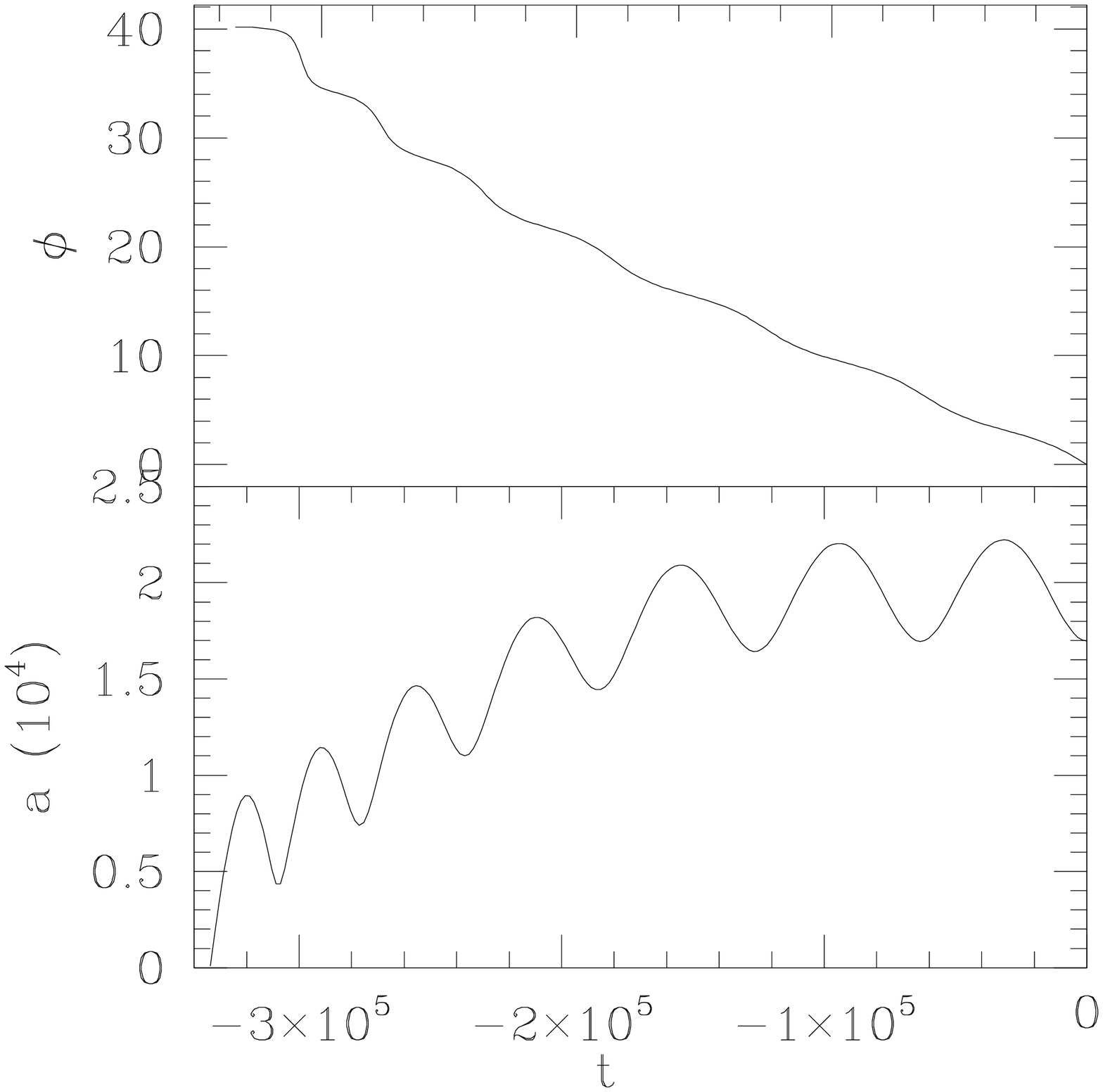}{Here alpha is taken as $10^{-13}$, with the rest of the
  parameters the same as in figure 1. The initial value of $\phi$ is carefully
  tuned to give the multiple oscillations for $a$}
  
  In the spirit of HH, we would interpret this universe as beginning in the
  Euclidean regime with $a=0$ and $\phi=i\Phi$ and $\dot\phi$ regular. The
  universe would "expand" and bounce between the potential and kinetic dominated
  bounce points for as long as desired, until it suddenly makes the transition
  to real $N$ and real $\phi$ at the lower kinetic dominated bounce point. The
  universe would then collapse into the future, until finally, either the
  potential dominated and the universe suffers a De Sitter, inflationary type
  bounce, or until the the universe finally collapses to a singularity. 
  
  \section{Sign of Euclidean Action}
  
  Since these solutions are possible, how would they enter into the path
  integral for the ``wave function of the universe"? There has of course been
  much heat expended on the various schools as to the sign one should take for
  the action in the Euclidean regime. These amount to either a choice of sign
  for $N$ in the imaginary regime, or a choice in the sign of $a$ (since $a$
  enters only quadratically in all expressions of physical relevance). Thus the
  contributions of these imaginary $N$ parts of the action will contribute
  either $e^S$ or $e^{-S}$ where $S$ is the integrated action for the imaginary
  $N$ regime (or is the imaginary part of the action in the case of a generic
  complex metric and field dominating the action). For our purposes, this will
  be the value of 
  \beq
  S_I= \int\left( -6a +a\dot a ^2  +a^3\left({1\over 2}\dot\Phi^2
  +V(I\Phi)\right)\right) dt
  \eeq
  During the bouncing, all the terms are of the same order as the first term,
  and we have
  \beq
  S_I\propto \int a dt
  \eeq
  By hypothesis both $a$ and $t$ are large, and furthermore, $t$ in general will be
  of the same order as $na$, where $n$ is the number of bounces in the imaginary
  $N$ regime, making $S_i$ of order of $na^2$.  $a$ here will be of order
  its maximum where $a^2\propto {1\over V0}$. Ie, the action will be of order
  ${n\over V_0}$. For small $V_0$, and for the positive choice of the $S$ in the
  action, this would completely dominate the path integral. The theory would
  predict with overwhelming probability that the universe began large and then
  collapsed down to a small size, rather than the more traditional way around.
  Furthermore, the fluctuations would grow during the collapse phase. The
  universe would grow more and more chaotic as its size decreased, with the
  simplest being when the universe was at its largest diameter. This sign is the
  sign usually attributed to Hartle and Hawking \cite{HH}. 
  
  The alternative sign, usually attributed to Linde and Vilenkin \cite{L}, \cite{V} (although
  derivable as simply one of two possibilities in defining the Hartle Hawking
  wave function) would on the other hand make the contributions of this
  particular solution to the semi-classical equations, completely sub-dominant.

  One of the interesting features of this analysis is the realization that the
  in order to maintain continuity of the scalar field equation  through the
  transition from the real to imaginary $N$, one must also allow $\phi$ to
  become complex. In particular, if the field and scale factor are real in the
  real $N$ regime, then the derivative of the field in general must be pure
  imaginary in the imaginary $N$ regime. Thus, we note that the sign of the
  kinetic term $a^3 {1\over 2N^2} \dot\phi^2$ does not change on the transition
  from real to imaginary $\phi$.
   
  \section{Fluctuations}
  Let us examine the action for the inhomogeneous minimally coupled scalar 
  field $\psi$ in the
  geometry. The action for this field is
  \beq
  S= \int Na^3\left( {1\over 2N^2}\dot\psi^2
      - {1\over a^2}(\partial\psi)^2\right)dt d\Omega^3
      \eeq
 where the $d\Omega^3$ is the spatial integral, and $(\partial\phi)^2$
 represents  the
 spatial derivative part of the action. Again at the surface where $N$ makes the
 transition from real to imaginary, and $\dot a$ is zero, $\psi$ must again be
 complex to keep $\dot\psi/N$ continuous. Ie, again we can write
 $\psi=\psi_0+i\Psi$. The action then becomes, taking $N=i$,
 \beq
 S_e= i\int  a^3({1\over 2} \dot\Psi^2 + {1\over a^2}(\partial \Psi)^2)
 dtd\Omega^3
 \eeq
 Taking the semi-classical approximation to the path integral with the VL sign
 we have  $e^{
 iS_e}$ as the contribution to the wave-function. All the terms in the action 
 are positive, indicating that we want to minimize the action to find the
 dominant terms. Using the apherical eigenstates for the spatial modes of the
 field we have the wave function factor as 
 \beq
 e^{ - \int a^3 ({1\over 2}\dot\Psi_k^2 + {k^2\over a^2} \Psi_k^2 )dt}
 \eeq
 near $a=0$, a will again be linear in $t$, and the solution for the field which
 is regular near $t=t_a$ is 
 \beq
 \Psi\approx \Psi_0(1 +{1\over 8} t^2 +...
 \eeq
 
 The action exponent can be written as
 \beq
 \int a^3 ({1\over 2}\dot\Psi_k^2 + {k^2\over a^2} \Psi_k^2
 )dt=a^3\Psi_k(t)\dot\Psi_k(t)\vert_{a=0}^t
 \eeq
 The lower limit is zero.
 As long as ${1\over a^{3\over 2}} {d^2 a^{3\over 2}\over dt^2}> {k^2\over a^2}$
 $\Psi_k$ will ``slow roll" and 
 \beq
 \dot\Psi_k\approx {k^2\over 3a\dot a}\Psi_k
 \eeq
 while in the other limit,
 \beq
 \dot\Psi_k\approx ({k\over a})\Psi_k
 \eeq

Thus, at the transition point, the exponent will be proportional to  
 $-\Psi_k^2$ (ie quadratic) and will be concentrated near zero with a width
 proportional to either $\sqrt{a\over k}$ for short wavelengths, or
 $\sqrt{H}\over k$ for long wavelengths.
 
 Thus, for the large initial conditions, the universe will start in its vacuum
 state at the largest turning point of the theory.

\end{document}